\documentstyle[aps,prl,preprint,epsfig]{revtex} 
\begin{document}                                     
\title{Structure and Instability of High-Density
Equations for Traffic Flow}
\author{Dirk Helbing}
\address{II. Institute for Theoretical Physics, University of
Stuttgart, 70550 Stuttgart, Germany}
\maketitle      
\draft
\begin{abstract}
Similar to the treatment of dense gases,
fluid-dynamic equations for the dynamics of congested vehicular traffic 
are derived from Enskog-like kinetic equations.
These contain additional terms due to the anisotropic 
vehicle interactions. 
The calculations are carried out up to Navier-Stokes order.
A linear instability analysis indicates an additional kind of 
instability compared to previous macroscopic traffic models.
The relevance for describing granular flows is outlined.
\end{abstract}
An efficient infrastructure is an essential precondition for every
industrialized country. Therefore, the considerable deterioration of the
traffic situation on `freeways' 
during the last decade is a serious problem.
Not only does impeded traffic cause economic losses of many billions of
dollars each year, it also produces serious ecological damages. Thus,
great efforts have been made to develop methods for traffic
optimization and forecasts, for which reliable traffic
simulations are necessary. Consequently, research on traffic 
dynamics has recently become a very important topic. 
During the last years, numerous results have been published on microscopic models
\cite{Micro,Cell1,MD,Hab}, 
including cellular automata models \cite{Cell1} 
and molecular-dynamics-like models
\cite{MD,Hab}, kinetic models \cite{Hab,Nav,Prig,Kin,Ref,Wag2,Klar}, and
macroscopic (fluid-dynamic-like) models
\cite{Macro,Hab,Nav,Prig,Kin,Ref,Wag2,Klar}, aiming at an understanding of
stop and go traffic. The topic is related to the fields 
of non-linear dynamics \cite{Macro}, 
phase transitions \cite{Micro,Cell1,MD,Hab,Macro}, and
stochastic processes \cite{Micro,Cell1}. 
\par
{\em Macroscopic} traffic models are not only suitable for on-line 
simulations of traffic networks, but also for analytical investigations.
Some very recent publications proposed to derive {\em realistic} traffic equations
from the `microscopic' dynamics of driver-vehicle
units via a kinetic approach \cite{Hab,Nav,Prig,Kin,Ref,Wag2,Klar}.
However, for the following reasons, these attempts have been only 
partly successful. Either the
models treat the vehicles like point-like objects \cite{Nav,Prig,Kin}. Then, the
resulting macroscopic models are only valid for free traffic flow.
Or the models take into account
the finite space requirements of vehicles \cite{Hab,Ref,Wag2,Klar}. However,
the calculations were carried out
up to {\em Euler} order only, based on the (zeroth order)
approximation of local equilibrium. This assumes that the {\em form} of the
equilibrium velocity distribution remains unchanged in dynamic situations, 
but it is given by the {\em local} values of the density $\rho(r,t)$,
average velocity $V(r,t)$, etc. 
It will be demonstrated that 
corresponding traffic models
are not even valid in linear approximation, since their linear
stability analysis gives totally misleading results.
To solve this problem, we must calculate the (first order) Navier-Stokes
corrections of the macroscopic traffic equations, which take into account the
structural change of the velocity distribution in inhomogeneous traffic
situations. 
\par
For ordinary gases, the Navier-Stokes terms (transport terms)
are calculated from the kinetic equation by means of the Chapman-Enskog method 
\cite{Lib}. An approximate, but more intuitive method 
bases on the relaxation time approximation \cite{Rel,Treiber} (see below).
For kinetic {\em traffic} models the situation is more involved,
since the interaction term does
not vanish in equilibrium situations.
Now, this problem has been solved. In the following, we
will show how to derive realistic traffic equations which
include corrections due to vehicular space requirements {\em and}
Navier-Stokes terms. Since these non-trivial 
corrections change the {\em structure} 
of the equations, they cause an additional instability. For the same reason, 
the low-density regime does not allow an extrapolation to situations 
at high densities. The presented method is also relevant for 
understanding instabilities in granular flows, 
since granular collisions are also not energy conserving. Recently,
a lot of publications tried to tackle these interesting problems with
fluid-dynamic equations derived from the associated
kinetic equation \cite{densegran,granfluids,pipefluid}. 
However, most of them are restricted to Euler order or the low-density regime 
\cite{granfluids,pipefluid}, thereby neglecting relevant sources of 
stability and instability. Probably for this reason, 
these approaches have not been fully successful in describing the 
formation of density waves in sand which is falling through a vertical pipe
\cite{pipefluid,graninst}. 
\par
{\it The model:} Since the number of vehicles 
on a (for simplicity: circular) freeway is conserved,
the kinetic traffic equation for the phase space density
$\tilde{\rho}(r,v,t)$ of vehicles with velocity $v = d_t r$ at place $r$ and time
$t$ has the form of a continuity equation with a sink/source term:
\begin{equation}
 \partial_t \tilde{\rho}
 + \partial_r (\tilde{\rho} v) 
 + \partial_v (\tilde{\rho} \; d_t v)
 =  ( \partial_t \tilde{\rho})_{\rm ss} \, .
\end{equation}
As usual, we will assume the acceleration law $d_t v = (V_0 - v)/\tau$,
where $\tau$ denotes a density-dependent acceleration time
and $V_0$ the desired velocity, which is assumed to be the 
same for all vehicles, here (case of a speed limit). The sink/source
term $(\partial_t \tilde{\rho})_{\rm ss}$ originates from sudden 
(non-differentiable) velocity changes. 
It splits up into a velocity-diffusion term due to
fluctuations of the acceleration behavior
(`imperfect driving') and an interaction term:
\begin{equation}
(\partial_t \tilde{\rho})_{\rm ss}
= \partial_v^2 (\tilde{\rho} D) + (\partial_t \tilde{\rho})_{\rm int} \, .
\end{equation}
The interaction term reflects sudden deceleration processes. In analogy to
the Enskog theory of dense gases \cite{chapman,densegas} and granular media
\cite{densegran}, but with an interaction law 
typical for vehicles \cite{Ref,Hab}, it is of the form
\begin{equation}
 (\partial_t \tilde{\rho})_{\rm int} = (1-p) \chi(r+l,t)
 {\cal B}(v) 
\end{equation}
with the Boltzmann-like interaction function
\begin{mathletters}
\begin{eqnarray}
 {\cal B}(v) &=& \!\!\int_{w>v}\!\!\!\!\! dw \, (w-v) \, \tilde{\rho}(r,w,t)
 \tilde{\rho}(r+s,v,t) \\
 &-& \!\!\int_{v>w}\!\!\!\!\! dw \, (v-w) 
 \tilde{\rho}(r,v,t)\tilde{\rho}(r+s,w,t) \, .
\end{eqnarray}
\end{mathletters}
According to this, the phase-space density $\tilde{\rho}(r,v,t)$ increases
due to deceleration of vehicles with velocities $w>v$, which cannot 
overtake vehicles with velocity $v$. The density-dependent
probability of immediate overtaking is represented by $p$.
A decrease of the phase space density $\tilde{\rho}(r,v,t)$ is caused by
interactions of vehicles with velocity $v$ with slower vehicles driving
with velocities $w<v$. The corresponding interaction rates are proportional
to the relative velocity $|v-w|$ and to the phase space densities 
of both interacting vehicles. By $s(V) = l_0 + l(V)$ ($\approx$ 
vehicle length $+$ safe distance) it is taken into account
that the distance of interacting vehicles is given by
their velocity-dependent space requirements.
These cause an increase of the interaction rate, which is described by the
pair correlation function $\chi(r) = [1- \rho(r,t) s]^{-1}$
at the 'interaction point' $r+l$.
A more detailled discussion of the above kinetic 
traffic model is presented elsewhere \cite{Hab,Ref}. 
By describing the individual acceleration behavior
via $d_t v=(V_0-v)/\tau$ and by introducing a velocity-diffusion term as 
source of velocity variations, it improves the original approach by 
Prigogine and Herman \cite{Hab,Nav}, which assumes 
a relaxation of the actual phase-space density to a desired one
\cite{Prig}. 
\par
Now, we will focus on the the macroscopic equations for the spatial density 
$\rho(r,t) = \int dv \, \tilde{\rho}(r,v,t)$, the average velocity
$V(r,t) = \int dv \, v \tilde{\rho}(r,v,t)/\rho(r,t)$, and the velocity variance
$\Theta(r,t) = \int dv \, [v-V(r,t)]^2 \tilde{\rho}(r,v,t)/\rho(r,t)$. These
are obtained by multiplying the kinetic equation with $v^k$, integrating
with respect to $v$, and a number of straight-forward calulations \cite{Hab,Ref}.
In order to underline the crucial results of this paper, we will first discuss
the case of {\em negligible} space requirements ($s, l, l_0 \ll 1/\rho(r,t)$), 
in which the macroscopic traffic equations read
\begin{eqnarray}
 \partial_t \rho + V \partial_r \rho &=& - \rho \, \partial_r V \, , \label{cont}
 \\
 \partial_t V + V \partial_r V &=& - 1/\rho \, \partial_r {\cal P}
 + (V_0 - V)/\tau - (1-p) {\cal P} \, , \label{vel} \\
 \partial_t \Theta + V \partial_r \Theta &=& - 2{\cal P}/\rho \, 
 \partial_r V - 1/\rho \,
 \partial_r {\cal J} + 2 (D -\Theta/\tau) \nonumber \\
 &-& (1-p) {\cal J} \, . \label{var}
\end{eqnarray}
Here, ${\cal P} = \rho \Theta$ denotes the `pressure' and
${\cal J}(r,t) = \rho(r,t)\Gamma(r,t)
= \int dv \, [v-V(r,t)]^3 \tilde{\rho}(r,v,t)$
the flow of velocity variance. (\ref{cont}) is the
expected continuity equation for the density. In comparison with the
conventional Euler equations for ordinary gases, the velocity equation
(\ref{vel}) and the variance equation (\ref{var}) contain two additional terms,
each of which breaks momentum and energy conservation. 
The respective last terms result from the anisotropic vehicle interactions,
while the previous terms reflect 
acceleration behavior and velocity fluctuations.
\par
It can be shown that the kinetic traffic equation
has the Gaussian equilibrium solution
$\tilde{\rho}_{0}(v) = \rho (2\pi\Theta)^{-1/2} \exp[-(v- V)^2/(2\Theta)]$,
which additionally fulfills the implicit equilibrium relations 
$V = V_0 - \tau (1-p) \rho \chi \Theta$ and $\Theta = D\tau$. 
If the local values $\rho(r,t)$, $V(r,t)$, and $\Theta(r,t)$ are inserted,
instead, we obtain the Euler approximation. 
It leads to ${\cal J}= \rho \Gamma \approx 0$ \cite{Hab,Fund}.
%
%
However, in {\em inhomogeneous} traffic situations the form of the
velocity distribution $P(v;r,t) = \tilde{\rho}(r,v,t)/\rho(r,t)$
changes due to the finite adaptation time $\tau_0$
which is needed to reach local equilibrium. In relaxation-time approximation 
\cite{Nav,Rel,Treiber} we find the Navier-Stokes correction 
$\Gamma = - 3\sqrt{\pi \Theta}/$$[\mbox{$(1-p)$}\rho]
\partial_r \Theta$. The corresponding instability diagram is 
depicted in Fig.~1 for the following model functions approximating
empirical results \cite{Hab,Ref}: 
$\tau(\rho) = 8\mbox{\,s} / [ 0.97\exp(-\rho/16\mbox{\,km}^{-1}) + 0.03]$,
$p(\rho) = \exp(-\rho/16\mbox{\,km}^{-1})$, and $D = 0.03 V^2/\tau(\rho)$.
\par
The instability diagram is obtained by (i) assuming a small
periodic perturbation
$\delta g(r,t) = g_0 \exp[ikr+(\lambda + {\rm i}\omega)t]$
of the macroscopic traffic quantities $g \in \{\rho, V, \Theta\}$
relative to the stationary and spatially homogeneous 
equilibrium solution $g_{\rm e}(\rho)$ ($g_0$ being the amplitude,
$k$ the wave number, $\lambda$ the growth rate, and $\omega$ the frequency
of the perturbation), 
(ii) inserting $g(r,t)= g_{\rm e} + \delta g(r,t)$
into the macroscopic traffic equations,
(iii) neglecting quadratic terms in the small perturbations
$\delta g (r,t) \ll g_{\rm e}$, (iv) determining the three
complex eigenvalues $\tilde{\lambda} = \lambda + {\rm i} \omega$
of the linearized equations in dependence of 
$\rho$ and 
$k$. Equilibrium traffic flow is unstable, giving rise to the well-known
phenomenon of {\it stop and go traffic} \cite{Hab,Macro}, if at least
one of the growth rates is positive, i.e. max $\lambda > 0$. Therefore,
the instability diagram shows max $\lambda(k,\rho)$ if this is greater
than zero, otherwise 0.
\par
It is interesting to compare Fig.~1 with the instability diagram 
in Fig.~2 which corresponds to the Euler approximation
with $\Gamma = 0$. This shows that the curious maxima in the middle of 
Fig.~1 are an effect of the deformation of the Gaussian velocity distribution 
in inhomogeneous situations. Since they originate from the terms containing
$\Gamma$, they are directly connected with the dynamic variance equations.
Therefore, they could not be discovered in 
previous traffic models which eliminated $\Theta(r,t)$ by means of approximations 
of the kind $\Theta(r,t) \approx \Theta^{\rm e}(\rho(r,t),V(r,t))$ \cite{Hab,Ref}.
\par
Apparently, the Navier-Stokes corrections do not cause a
stability of equilibrium traffic flow with respect to perturbations of large
wave numbers $k$ (i.e.\ small wave lengths $\ell = 2\pi/k$). This
surprising result is a serious problem for a numerical solution 
of the above equations. It comes from the fact that
(shear) viscosity terms are missing 
due to the spatial one-dimensionality of traffic flow.
This problem vanishes when corrections due to vehicular space requirements
are taken into account ($l_0 = 1 / \rho_{\rm max}$, $l = 0.8\mbox{\,s}\cdot V$).
One would not expect this, since, for ordinary
gases, the {\em structure} of the fluid-dynamic equations does not change at high
densities. The only thing what changes are the constitutive 
relations for ${\cal P}$ and ${\cal J}$ \cite{chapman,Hab}. 
However, in the case of traffic dynamics the situation is completely
different. Since the anisotropic vehicle interactions do not fulfil momentum and
energy conservation, they lead to contributions that cannot be
absorbed by modified functions ${\cal P}$ and ${\cal J}$. Whereas the
vehicle density still obeys the continuity equation (\ref{cont}),
the structure of the velocity and variance equations 
changes considerably ($\gamma = (1-p) \chi$):
\begin{eqnarray}
 \partial_t V + V \partial_r V &=& -[1/\rho + \gamma s 
 (1 + \rho \chi l) ]\Theta \, \partial_r \rho \nonumber \\
 &+& \gamma \rho (2s\sqrt{\Theta/\pi} - \rho \Theta \chi l^2/V) \,
 \partial_r V \nonumber \\
 &-& [1 + \gamma\rho s/2] \, \partial_r \Theta \nonumber \\
 &-& \gamma s (s/2 + \rho \chi l^2/2) \Theta \, \partial_r^2 \rho
 \nonumber \\
 &+& \eta / \rho  \, \partial_r^2 V 
 - \gamma \rho s^2/4 \, \partial_r^2 \Theta \nonumber \\
 &+& (V_0 - V)/\tau - \gamma \rho \Theta \label{newvel} \\
 \partial_t \Theta + V \partial_r \Theta &=& 
 - [2 + \gamma\rho s]\Theta  \, \partial_r V \nonumber \\
 &+& 2\gamma \rho s \sqrt{\Theta/\pi} \, \partial_r \Theta 
 - \gamma \rho  s^2 \Theta/2 \, \partial_r^2 V \nonumber \\
 &+& \gamma \rho s^2 \sqrt{\Theta/\pi} \, \partial_r^2 \Theta 
 + 2(D - \Theta/\tau) \, . \label{newvar}
\end{eqnarray}
This result is valid up to Euler order. It has been
obtained by evaluating the kinetic equation on the assumption of 
a Gaussian velocity distribution $P_{0}(v;r,t)$ and second order
Taylor expansion of the functions $\tilde{\rho}$ and $\chi$ 
with respect to $s$ and $l$ around $r$, thereby neglecting products of
partial derivatives. Fig.~3 shows that the finite
space requirements of vehicles cause the desired stability of equilibrium
traffic flow with respect to perturbations of small wave lengths. This comes
from the finite viscosity coefficient $\eta = 
\gamma \rho^2 [s^2 \sqrt{\Theta/\pi} - \rho\Theta\chi l^3/(2V) ]$.
\par
Nevertheless, the result is not even correct in linear approximation, 
since inhomogeneous traffic again changes 
the form of the velocity distribution. 
To calculate the Navier-Stokes corrections, 
we must derive an equation for
the small deviation $\delta \tilde{\rho}(r,v,t)$ from $\tilde{\rho}_{0}
  (r,v,t) = \rho(r,t)P_{0}(v;r,t)$ which is
caused by inhomogeneities $\partial_r \rho$,
$\partial_r V$, and $\partial_r \Theta$. This has been done by means of
the relaxation time approximation \cite{Nav,Rel,Treiber}
which assumes that 
(i) the deformation of the local equilibrium distribution
$P_0(v;r,t)$ is caused by the interaction term,
(ii) the non-equilibrium corrections of the latter can be
adiabatically approximated by $- \delta \tilde{\rho}(r,v,t)/\tau_0$,
where $-1/\tau_0$ denotes the slowest eigenvalue of the linearized interaction
operator, (iii) $1/\tau_0$ is of the order of the vehicular interaction rate
\begin{eqnarray}
 \frac{1}{\tau_0(r,t)} &=& (1-p)\chi(r+l,t)/\rho(r,t) \nonumber \\
 &\times & \!\!\int\!\! dv \!\!\int_{w<v}\!\!\!\!\!
 dw \, |v-w| \tilde{\rho}(r,v,t) \tilde{\rho}(r+s,w,t) \, .
\end{eqnarray}
The finally resulting relation is, with $\delta v = v-V$,
\begin{eqnarray}
 \delta \tilde{\rho}(r,v,t) &\approx &  \tilde{\rho}_{0}(r,v,t) 
 \tau_0( 3\, \delta v/\Theta  - \delta v^3/\Theta^2 )/2
 \, \partial_r \Theta  \nonumber \\
 &+& \tau_0 (1-p) \chi(r+l,t) {\cal B}(v) \nonumber \\
 &-& P_{0}(r,v,t) \tau_0 \delta v/\Theta (1-p)
 \chi(r+l,t) \!\!\int\!\! dv \, \delta v {\cal B}(v) \nonumber \\
 &-& P_{0}(r,v,t) \tau_0/(2\Theta^2) (\delta v^2 -
 \Theta) (1-p) \chi(r+l,t) \nonumber \\
 &\times & \!\!\int\!\! dv \, \delta v^2 {\cal B}(v) \, .
\end{eqnarray}
With $\tilde{\rho}(r,v,t) \approx \rho_{0}(r,v,t) + \delta
\tilde{\rho}(r,v,t)$, it follows 
that the quantities $\rho(r,t)$,
$V(r,t)$, and $\Theta(r,t)$, which are taken into account by the
Gaussian approximation $P_0(v;r,t)$, are not corrected by
$\delta \tilde{\rho}(r,v,t)$.
However, for the third central velocity moment $\Gamma$ we obtain,
instead of $\Gamma \approx 0$, 
\begin{eqnarray}
 \Gamma &=& - 3\sqrt{\pi\Theta}/[(1-p)\rho\chi] \, \partial_r \Theta
 + s\Theta \, \partial_r V + s^2 \Theta/2 \, \partial_r^2 V \nonumber \\
 &-& 3 s \sqrt{\pi\Theta}/2 \, \partial_r \Theta - 3 s^2 \sqrt{\pi\Theta}/4 \,
 \partial_r^2 \Theta \, ,
\end{eqnarray}
which becomes different from zero in inhomogeneous traffic situations.
This causes the additional contribution
\begin{equation}
 + \rho s / (6\sqrt{\pi\Theta}) \, \partial_r \Gamma + \rho s^2 /
 (12\sqrt{\pi\Theta}) \, \partial_r^2 \Gamma
\end{equation}
to the velocity equation (\ref{newvel}) and the extra term
\begin{eqnarray}
 &-& \Gamma [\rho + s \, \partial_r \rho + s^2/2 \, \partial_r^2 \rho
 - \rho s /\sqrt{\pi \Theta} \, \partial_r V \nonumber \\
 & & - \rho s^2/(2 \sqrt{\pi\Theta})
 \, \partial_r^2 V ] - \rho s /2 \, \partial_r \Gamma - \rho s^2/4
 \, \partial_r^2 \Gamma
\end{eqnarray}
to the variance equation (\ref{newvar}). Together with the continuity equation
(\ref{cont}), the resulting equations are
the desired macroscopic traffic equations
for high densities. The related instability diagram is
depicted in Fig.~4 and indicates two different kinds of
instabilities.
\par 
In summary, we have found several significant results, which are
not sensitive to the particular choice of the parameters or 
to different variations of the model:
(i) A conistent traffic model needs to take into account vehicular
space requirements as well as Navier-Stokes terms in order to allow a
realistic description of traffic instabilities.
(ii) Treating vehicles in a point-like manner, the stability of
equilibrium traffic flow with respect to perturbations of small wave lengths
is not correctly described, even if Navier-Stokes terms are included.
(iii) This problem vanishes when corrections due to the finite space
requirements of vehicles are considered. 
That is, the macroscopic traffic equations for low densities do not
allow an extrapolation to the traffic dynamics at high densities.
(iv) The Navier-Stokes terms are responsible for a subdivision of the
instability region into separate areas. These belong to different
eigenvalues. 
(v) Whereas the instability diagrams of traffic models,
which consist of a density and a velocity equation only,
typically show two relevant 
humps \cite{Hab,Ref}, the two additional (narrow) humps of the
above model are related to the dynamic variance equation. Therefore, 
the dynamic variance equation gives rise to 
a new kind of traffic instability. 
\par
A more detailled discussion of the applied relaxation time approximation
as well as of equations, simulations, and results for the non-linear regime
of traffic dynamics will be given in a forth-coming paper \cite{Treiber}.

\clearpage
{\small \unitlength1cm
\epsfig{width=8.5\unitlength, file=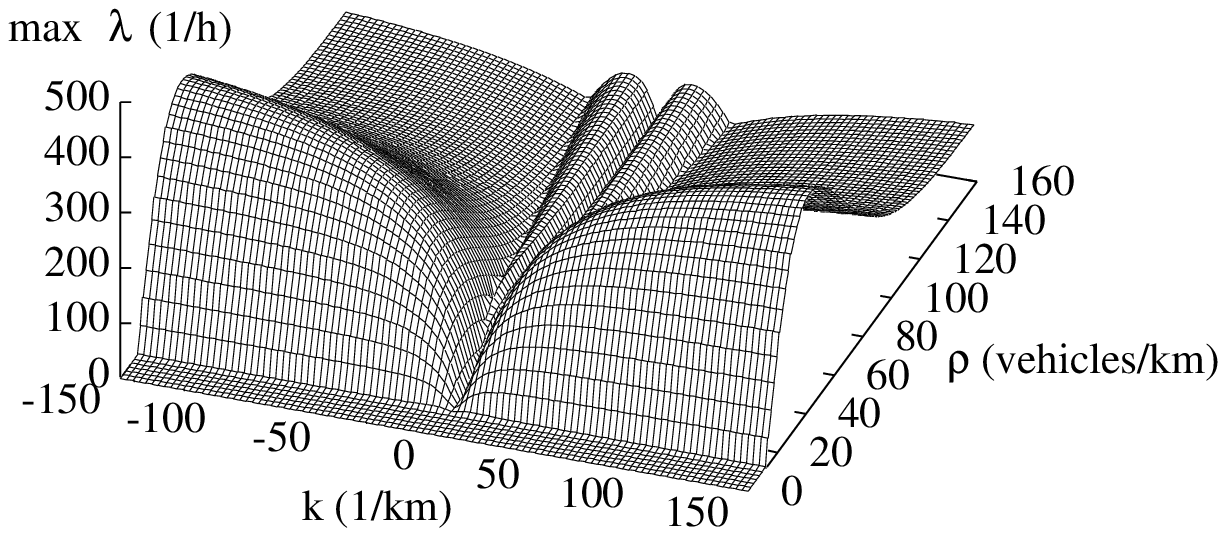}
Fig.~1: The instability diagram 
of the Navier-Stokes-like traffic model 
for point-like vehicles ($s = l = l_0 = 0$) indicates that traffic flow would be
unstable above a certain critical density,
surprisingly even at large wave numbers $k$.

\epsfig{width=8.5\unitlength, file=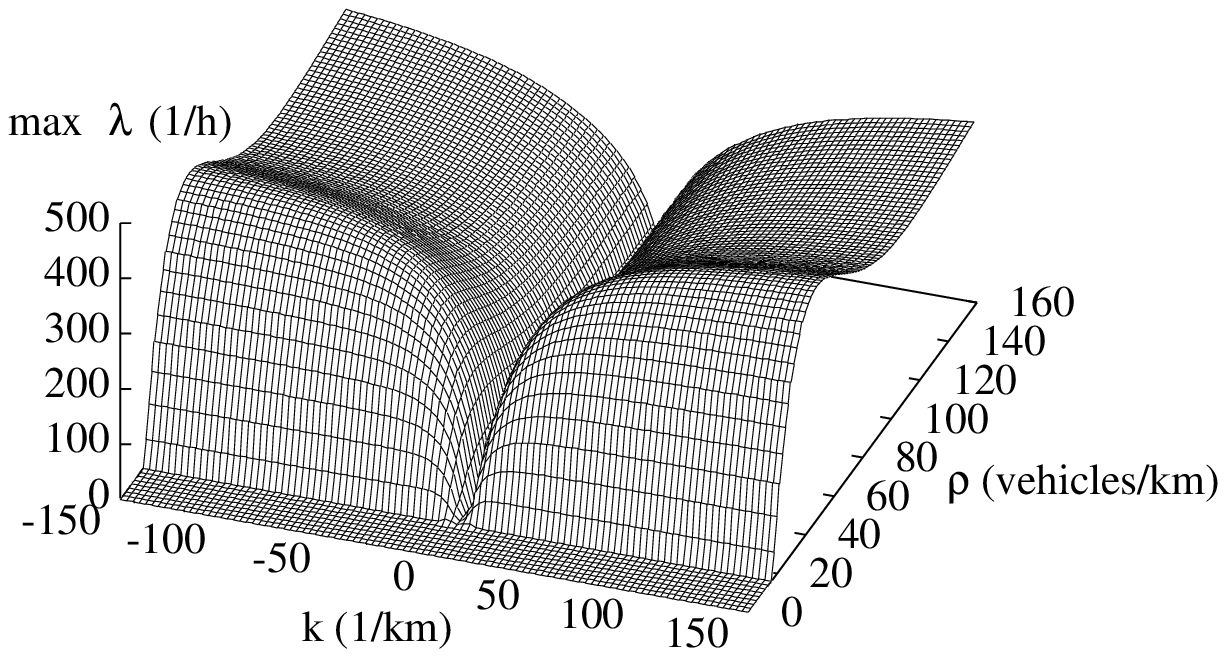}
Fig.~2: Same as Fig.~1, but for macroscopic traffic equations calculated
up to Euler order only, thereby neglecting the deformation of 
the velocity distribution in inhomogeneous traffic situations.
\clearpage
\epsfig{width=8.5\unitlength, file=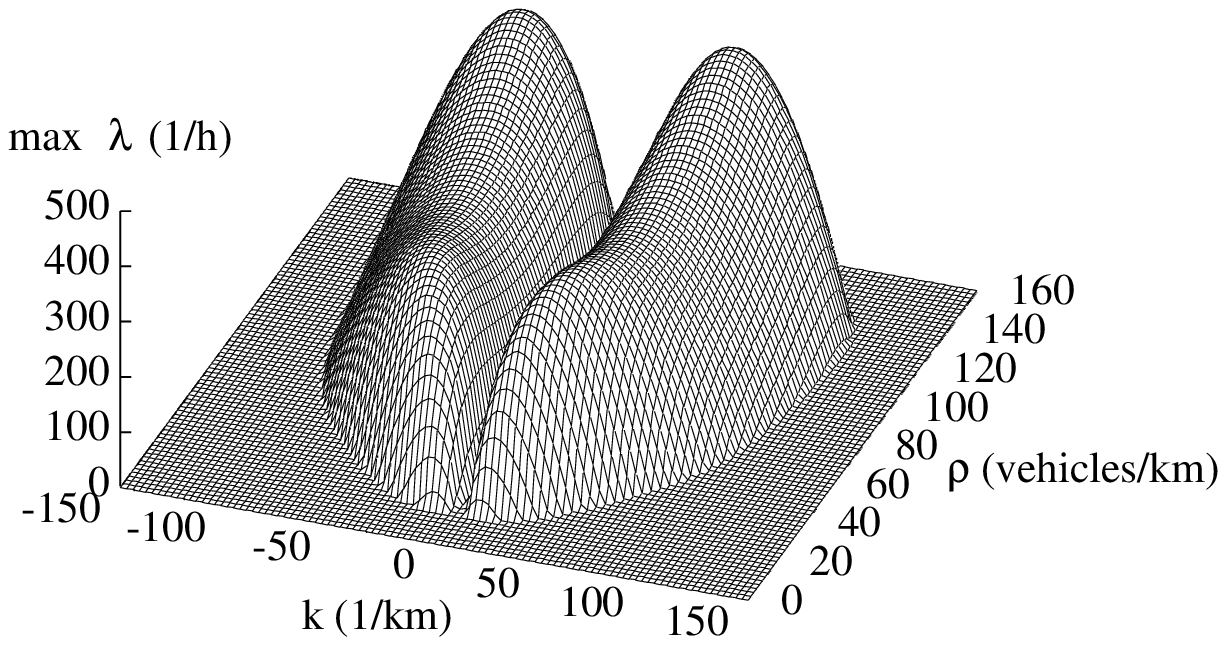}
Fig.~3: Same as Fig.~2, but with consideration of
vehicular space requirements ($l_0 = 1/\rho_{\rm max}$,
$l = 0.8\mbox{\,s}\cdot V$). In agreement with empirical findings \cite{Fund},
the results predict that equilibrium traffic flow is stable up to
12 vehicles per kilometer and lane, at extreme
densities, and at high wave numbers (small wave lengths). 

\epsfig{width=8.5\unitlength, file=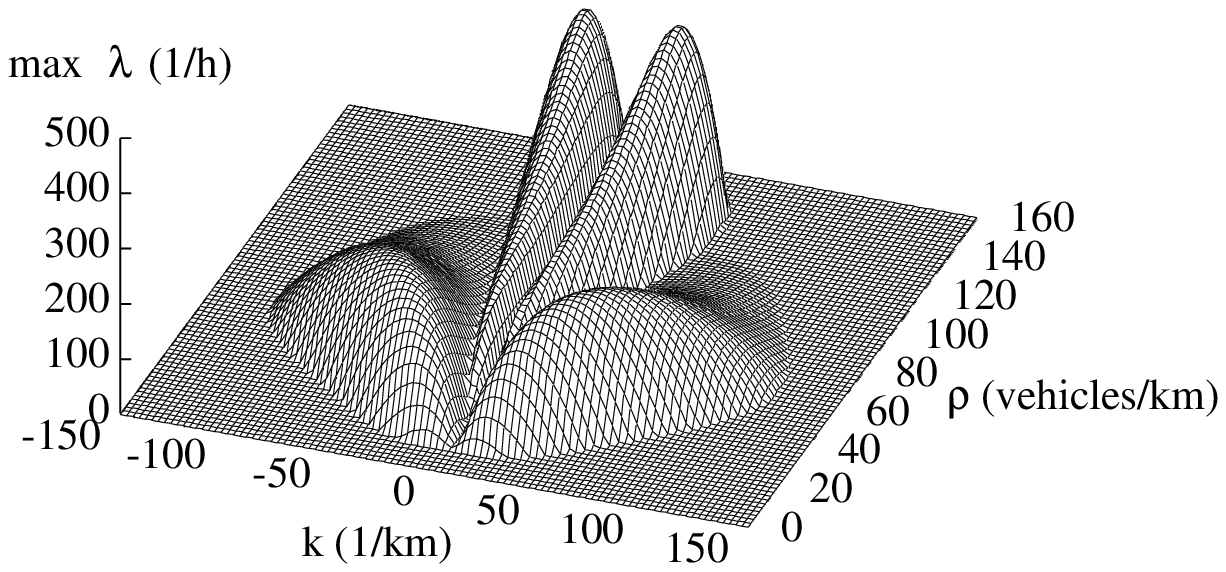}
Fig.~4: Same as Fig.~3, but for macroscopic traffic equations including
vehicular space requirement {\em and} Navier-Stokes corrections in order to
obtain valid results. The instability diagram is now 
divided into two separate humps in each half-plane, indicating two
different kinds of instability. 
}
\end{document}